\documentstyle[12pt, aaspp4]{article}

\def\ltsima{$\; \buildrel < \over \sim \;$}
\def\simlt{\lower.5ex\hbox{\ltsima}}
\def\gtsima{$\; \buildrel > \over \sim \;$}
\def\simgt{\lower.5ex\hbox{\gtsima}}

\begin{document}

\title{The Spin Period, Luminosity and Age Distributions of Anomalous 
X-Ray Pulsars}

\author{Pinaki Chatterjee$^1$ \& Lars Hernquist$^2$}
\affil{Harvard-Smithsonian Center for Astrophysics, 60 Garden
Street, Cambridge, MA 02138}
\footnotetext[1]
{pchatterjee@cfa.harvard.edu}
\footnotetext[2]
{lars@cfa.harvard.edu}

\medskip

\begin{abstract}

We consider the accretion model for anomalous X-ray pulsars proposed
recently by Chatterjee, Hernquist and Narayan, in which the emission is
powered by accretion from a fossil disk formed by the
fallback of material from a supernova explosion. We demonstrate that
this model is able to account for the spin period, luminosity and age
distributions of the observed population of AXPs for reasonable and
broad distributions of the free parameters of the model, namely, the
surface magnetic field of the neutron star, the mass of its accretion
disk and its initial spin period. In particular, this model is able
statistically to account for the puzzlingly narrow observed spin 
distribution of the AXPs. We show also that if the establishment of  
fallback accretion disks around isolated neutron stars is a universal
phenomenon, then a fairly large minority ($\sim 20\%$) of these objects
become X-ray bright AXPs or X-ray faint systems spinning down by 
propeller action, while the rest become radio pulsars.

\end{abstract}


\section{Introduction}

Anomalous X-ray pulsars (AXPs), about half a dozen of which are known,
have properties significantly different from those of binary X-ray
pulsars (see Mereghetti 1999 for a recent review).  AXPs are sources
of pulsed X-ray emission with relatively low persistent X-ray
luminosities, $L_x \sim 10^{35} - 
10^{36}$ erg/sec, and soft spectra which are well-fitted by a 
combination of blackbody and power-law contributions, with effective 
temperatures and photon indices in the range $T_e \sim 0.3 - 0.4$ keV 
and $\Gamma \sim 3 - 4$, respectively. They have relatively long spin
periods of about $P\sim 6 - 12$ seconds, which increase steadily with
time. Their characteristic ages are about $P/2\dot{P} \sim 10^{3} - 
10^{5}$ years. No binary companions have been detected for these
objects, and observations have placed strong constraints on companion
masses (e.g. Mereghetti, Israel \& Stella 1998; Wilson et
al. 1998). At least three AXPs have been associated with young
supernova remnants which limit their ages to $\sim 10 - 20$ kyr. 

There are at present two theories seeking to explain the properties of
AXPs. In one, they are modeled as isolated, ultramagnetized neutron
stars or ``magnetars'' spinning down by the emission of magnetic
dipole radiation, which is powered either by residual thermal energy
(Heyl \& Hernquist 1997a,b) or by magnetic field decay (Thompson \& 
Duncan 1996). The required surface magnetic field strength is $B \sim 
10^{14} - 10^{15}$ G, which is similar to the values inferred from 
timing data for soft gamma repeaters or SGRs (e.g. Kouveliotou et al. 
1998, 1999; see, however, Marsden, Rothschild \& Lingenfelter 1999). 

The other class of theories proposes that the X-ray emission from AXPs
is powered by accretion from binary companions of very low mass 
(Mereghetti \& Stella 1995), from the interstellar medium (see Wang
1997 in reference to the AXP candidate RX J0720.4-3125; however, see 
Heyl \& Hernquist 1998 for a magnetar model of this object), or from
the debris of a disrupted high-mass X-ray binary system, after a 
stage of common-envelope evolution (van Paradijs et al. 1995; Ghosh, 
Angelini \& White 1997). The inferred magnetic field strengths of the 
neutron stars then are similar to those of ordinary radio pulsars and 
luminous X-ray pulsars ($B \simlt 10^{12}$ G).

More recently, Chatterjee, Hernquist and Narayan (1999; hereafter CHN)
proposed another accretion model in which AXPs are neutron stars
with standard magnetic fields accreting from a disk formed after
fallback of material from a supernova explosion. The star is rapidly 
spun down close to the observed AXP periods on a timescale of $\sim
10^{4}$ years by the action of a propeller effect (for other accretion 
scenarios for AXPs and SGRs based on the propeller effect, see Alpar
[1999] and Marsden et al. [1999]). 

However, it has been observed that the spin-down of certain AXPs is
very stable over time (Kaspi, Chakrabarty \& Steinberger 1999). This
could pose a problem for all accretion models, in which a higher level
of timing noise might be expected. Moreover, it is not clear whether
optical and infra-red emission from such accretion disks would be low
enough to evade the limits set by observations (Perna, Hernquist \&
Narayan, 1999; Hulleman et al. 2000). It remains to be seen whether 
accretion models can circumvent these difficulties.

A question that has hitherto largely been unaddressed is the reason
for the puzzlingly strong clustering of spin periods of the observed 
AXPs. Colpi, Geppert \& Page (2000) contend that this phenomenon implies
magnetic field decay in the magnetar model. Marsden et al. (1999) seek 
to explain this by limiting the magnetic fields of the neutron stars 
and the strength of the propeller wind emission by which they spin
down to narrow ranges of values. In this paper, we show that the
observed narrow period range of AXPs is consistent with the accretion 
model described in CHN if the neutron stars are drawn from an
underlying population characterized by broad and reasonable
distributions of magnetic field strength, initial spin and accretion 
disk mass.

It has been suggested that AXPs and SGRs are drawn from the
same underlying population of objects. This is motivated by the
observation that SGRs (see Hurley 1999 for a review) have properties 
that are extremely similar to those of AXPs, except for the fact that 
they occasionally undergo energetic outbursts. We show in this paper that
the narrow period range of AXPs and SGRs taken together is also
consistent with the above model. 

\section{Summary of Possible Neutron Star Spin Histories}

The theoretical model used in this paper is that which is set out in 
detail in CHN, and to which the reader is referred for details. Here
we only summarize the different possible histories of a neutron star
accreting from a debris disk. 

Such a star will have various spin-down histories depending on the
initial values of its free parameters: its surface magnetic field 
strength $B$, its initial spin period $P_{0}$, and the mass of its
accretion disk $M_d$. For some values of initial parameters, the star
goes into an extended radio pulsar phase in which the magnetospheric
radius is greater than the light cylinder radius; accretion action is
then assumed to stop. For other initial values, the star is spun down 
in the propeller phase, during which accreted matter is flung out by
centrifugal forces prior to reaching the surface of the star; the star
then reaches the quasi-equilibrium ``tracking phase'', in which the
spin of the star roughly matches the equilibrium period but never
quite equals it since the mass accretion rate declines with time. If
the star enters the tracking phase at a time $t_{trans}$, then we
identify the period between $t_{trans}$ and the ``accretion death
time'' $2 t_{ADAF}$ to be the AXP phase; here, $t_{ADAF}$ is the time
at which the mass accretion rate falls to about 1\% of the Eddington
value and the infall becomes an advection-dominated accretion flow or 
ADAF. For yet other initial values of the star's parameters, it
remains in the propeller phase until after the time $2 t_{ADAF}$; in 
other words, $t_{trans} > 2 t_{ADAF}$. Such a system was called 
a ``propeller system'' in CHN. 

The X-ray luminosity of the system is determined by the mass accretion
rate on to the surface of the star, $\dot{M_X}$, which could be
different from the mass accretion rate through the disk, $\dot{M}$, if
some of the material is driven away from the system before reaching
the surface of the neutron star. Thus, when accretion action ceases
and the star is in the radio pulsar phase, the X-ray
luminosity falls to zero. During the propeller phase, most of the 
matter will be ejected by centrifugal forces, and
the system will be X-ray faint, since $\dot{M_X} \ll \dot{M}$. During
the tracking phase, we assume that $\dot{M_X} \sim \dot{M}$ if
$\dot{M} \leq \dot{M_E}$, where $\dot{M_E}$ is the mass accretion rate
corresponding to the Eddington luminosity, but that $\dot{M_X} = 
\dot{M_E}$ if $\dot{M} > \dot{M_E}$; in this phase, the system will 
be X-ray bright. Note that the assumption here is that even though 
$\dot{M}$ could be highly super-Eddington, especially at early times, 
at no point would the X-ray luminosity exceed the Eddington limit,
owing to ejection of matter by radiation pressure prior to reaching 
the surface of the star. 

It is not known how rapidly the luminosity will fall during the ADAF
phase of the AXP; presumably it will do so faster than linearly in 
$\dot{M}$; in this paper, for purposes of numerical calculation, we 
assume the dependence of luminosity between between times $t_{ADAF}$ 
and $2 t_{ADAF}$ to be $L_X \propto \dot{M_X}^{2}$ (of course, for 
$t < t_{ADAF}$, $L_X \propto \dot{M_X}$; and for $t > 2 t_{ADAF}$, 
$L_X = 0$).

\section{The Distribution of Spin Periods of AXPs}

There are 6 AXPs known today: 1E 2259+586, RXS J170849-4009, 4U
0142+615, 1E 1048-5937, 1E 1841-045 and AXJ 1845-0300 (see Mereghetti
1999). Their spin periods, which are very tightly clustered together,
are, in seconds, 6.98, 11.00, 8.69, 6.45, 11.76 and 6.97,
respectively. Associated young supernova remnants (SNRs) of estimated
age $\sim 10^4$ years have so far been detected for 3 of these 6
objects. If it is assumed, as in the model described in the previous
section, that all the 6 observed AXPs are the result of supernova
explosions $\sim 10^4$ years ago, and that the birthrate of radio
pulsars in the galaxy is approximately 1 every 100-200 years 
(see, e.g., Lyne et al. 1998), then the relative birthrates of AXPs 
and radio pulsars turns out to be $\sim 0.06 - 0.12$. 

How is the distribution of spin periods of the AXPs to be calculated?
The evolution in time of the properties of an AXP depends on the three
free parameters of the model, namely, the surface magnetic field
strength of the neutron star $B_{12}$, the mass of the initial
circumstellar accretion disk $M_d$, and the initial spin period of the
star $P_0 (\equiv 2 \pi/ \Omega (0))$. The way we calculate the
distribution of AXP properties is as follows: we specify intrinsic
prior probability distributions for each of the parameters $B_{12},
M_d$ and $P_0$, and endow each neutron star with initial values of
$B_{12}, M_d$ and $P_0$ sampled randomly and independently from their 
respective distributions, as well as a time of birth chosen randomly 
from a distribution uniform in time, since we expect the birthrate of 
neutron stars to be constant over time. Having done this for a large 
number of model neutron stars, we evolve each until a later fiducial 
time, when we take account of how many stars at that fiducial time are 
bright AXPs, dead AXPs (i.e., an AXP which is past its $2 t_{ADAF}$ time),
propeller systems or radio pulsars. The number of sample neutron stars
should be sufficiently large, and the width of their time-of-birth
distribution sufficiently long, that the results of this calculation
are stable across repetitions of the process.

As an example, we present in detail the results of one such
calculation. We take the surface magnetic fields of the neutron stars
to be the distribution reported in Narayan \& Ostriker (1990) for radio
pulsars: namely, that Log $B$ is distributed normally with a mean of
12.5 and a standard deviation of 0.4 (note that in the rest of this
paper, all indicated logarithms are to be taken to base 10). The 
distributions of $P_0$ and $M_d$ are poorly constrained by
observations. We choose distributions for them that are simple and
cover fairly broad ranges of values. $P_0$ is chosen to be
distributed uniformly between 2ms and 50ms, and Log$(M_d/M_{\odot})$ 
is assumed to be distributed uniformly between -6 and -1.8.

In one particular realization of this calculation, the initial number
of neutron stars was selected to be 5,000,000; at the chosen fiducial
time, the number of bright AXPs was 139,306; bar graphs showing their
distribution of period, luminosity and age are shown in Figure 1. The
fractions of all initial neutron stars becoming AXPs, propeller
systems and radio pulsars were 0.07, 0.10, and 0.83,
respectively. Thus, the relative birthrates of AXPs and radio pulsars 
was 0.09. It will be observed that the AXP spin period distribution 
peaks at roughly 5-6 seconds, that the ages of the AXPs are of the order 
of a few times $10^4$ years, and that the X-ray luminosity
distribution peaks at about $4 \times 10^{35}$ ergs/s; all of these 
properties correspond reasonably well with the properties of observed 
AXPs.

Notice also that the spin period distribution has a long tail towards 
high periods. However, given that the total number of observed AXPs
(whose periods are much more closely clustered) is only 6, they could 
still constitute a representative sample from the above
distribution. To quantify this statement, we carry out the
nonparametric Kolmogorov-Smirnov (K-S) test for two discrete samples
(see, e.g., Press et al. 1988) to test whether the hypothetical and 
observed AXP spin period distributions are compatible.

We define the test statistic to be the quantity $S_{mn}
\equiv \big( \frac{mn}{m+n} \big) ^{1/2} D$, where $m$ and $n$
are the sizes of the observed and hypothetical spin period
distributions, respectively (here, $m=6$ and $n=139,306$), and $D$
is the maximum absolute difference between the cumulative distribution
functions of the above distributions. In the above 
realization of the calculation, $S_{mn}$ has the value 0.76, and it 
is just significant at the level of significance 0.61. The
cumulative distributions of the observed and hypothetical spin periods 
are shown in Figure 2.

We thus conclude that the observed AXP period distribution is
consistent with the hypothetical distribution predicted by our model.

The strongest factor determining the AXP properties in our model
is the $B$ distribution; the spin period distribution in Figure 1 can
be reasonably well-fitted by a log-normal distribution, reflecting the
log-normal distribution of $B$. The weakest factor is the initial spin
period $P_0$; changing its distribution does not significantly alter
the final AXP properties.

The above results are for the case in which Log$(M_d/M_\odot)$ is
distributed uniformly between -6 and -1.8. The AXPs in this model come
from the relatively high end of the $M_d$ distribution (in
the above realization, the peak value for the AXPs of 
Log$(M_d/M_\odot)$ is -1.8, and the minimum is at approximately -3.4);
neutron stars with smaller disk masses end up as propeller systems or
radio pulsars.  

One would thus expect that lowering the lower limit of the 
Log$(M_d/M_\odot)$ distribution would not change the AXP properties
described above, except that it would lower the ratio of birthrates 
of AXPs and radio pulsars, since a greater fraction of neutron stars 
would then have accretion disks too little massive to become
AXPs. This is indeed the case: for lower limits of -5, -6 and -7, the 
corresponding fractions of neutron stars which become AXPs are 0.09, 0.07 
and 0.06; the fractions of neutron stars becoming X-ray dim
propeller systems are 0.14, 0.10 and 0.08, respectively; the remaining 
neutron stars become radio pulsars, the fractions of which are 0.77, 0.82
and 0.86, respectively; hence, the corresponding ratios of birthrates
of AXPs and radio pulsars are 0.12, 0.09 and 0.07. 

On the other hand, changing the upper limit of the Log$(M_d/M_\odot)$
distribution changes not only the AXP-radio pulsar birthrate ratio,
but also the AXP spin distribution: raising it shifts the peak of the
period distribution towards lower periods, and lowering it shifts the
peak towards higher periods. This affects substantially the level of
significance of the K-S test: for upper limits of -1.7, -1.8 and -1.9,
the values of $S_{mn}$ are approximately 0.90, 0.76 and 0.89
respectively, with corresponding levels of significance of 0.39, 0.61 
and 0.41. The relative birthrates of AXPs, propeller systems and radio
pulsars also change: the respective fractions of neutron stars
becoming AXPs are 0.09, 0.07 and 0.06; the fractions of neutron stars
which become propeller systems are 0.11, 0.10 and 0.10, repsectively;
the corresponding fractions of neutron stars which become radio
pulsars are 0.81, 0.83 and 0.84; hence, the ratios of birthrates of
AXPs and radio pulsars for the above three cases are 0.11, 0.09 and
0.07, respectively. 

It is interesting to apply our model to try to explain the narrow
period distribution of the AXPs and SGRs taken together. There are
four known SGRs: SGR 1627-41, SGR 0525-66, SGR 1806-20 and SGR
1900+14; their spin periods are, in seconds, 6.41, 8.0, 7.48 and 5.16
respectively. If we take the 10 AXPs and SGRs together and apply the
K-S test to check whether this distribution is consistent with our
usual model with  Log$(M_d/M_\odot)$ distributed uniformly between -6
and -1.8, we find (with $m$ = 10, in this case) that $S_{mn} = 0.98$,
and it is just significant at the level of significance 0.29. Thus,
the combined AXP and SGR spin distribution is also consistent with our
hypothetical distribution. 

\section{Conclusion}

We have shown in this paper that according to our model for fallback
accretion disks around solitary neutron stars, not all isolated
neutron stars which are the product of supernova explosions become
radio pulsars; a substantial fraction of them ($\sim 5-10\%$, for the
models considered here) end up as X-ray  bright AXPs, a possibly
larger fraction ($\sim 8-14\%$) as X-ray faint propeller systems,
while the remainder end up as radio pulsars. If the 6 observed AXPs
were produced by supernova explosions $\sim 10^4$ years ago, then the
ratio of birthrates of AXPs and radio pulsars is expected to be of the
order of 10\%.

Whether a neutron star becomes a radio pulsar, an AXP or a propeller
system depends on the values of its surface magnetic field $B$,
initial spin period $P_0$, and initial circumstellar disk mass $M_d$. 

Using the observationally constrained distribution function of $B$ for
radio pulsars, and assuming simple distribution functions of initial
spin period and disk mass covering a broad range in $P_0$ and
Log$(M_d)$ respectively, it has been shown that our accretion model 
produces a population of AXPs with a spin period distribution peaking 
at $\sim 6$ seconds, X-ray luminosity $\sim 10^{35}$ ergs/s, age
$\sim 10^4$ years, and with approximately the above AXP-radio pulsar
birthrate ratio. Systems with relatively high values of $B$ and $M_d$
end up as AXPs, while those with low values end up as radio pulsars;
systems with intermediate values of these parameters become propeller
systems.

It is difficult to make definitive statements about possible relations
between the properties of the underlying neutron star, such as its
magnetic field strength or its initial accretion disk mass, and its
observable properties such as spin period, age or luminosity,
since the wide ranges of values the initial parameters can take give
rise to a lot of scatter in the final AXP properties that can be
observed. Nevertheless, some rough general trends may be discerned. 
For example, if a neutron star has a low disk mass (``low'' and ``high''
here are to be understood relatively in the context of the
corresponding distributions considered in the previous section and in
Figure 1), then it would need a high surface magnetic field strength 
in order to become a visible AXP; this would in turn tend to produce a 
probable observed spin period which lies slightly rightwards of the
peak of the spin period distribution in Figure 1; the probable
observed age is low, and the observed luminosity could take a range of 
values. On the other hand, a star with a high disk mass could reach
the AXP phase with a range of surface magnetic field strengths, and
the probable observed spin period, luminosity and age could also span a
range of values.  

In this paper, we wished also to address ourselves to the narrow
spin period range observed in AXPs. One might suspect that
this would in turn constrain the parameters of any model of AXPs to 
unphysically slender ranges of values. However, we have shown in this
paper that even under the assumption of quite reasonable and broad
distributions of the parameters, our model leads to a spin period 
distribution that is statistically consistent with the narrowly
clustered distribution observed in the case of the  
6 known AXPs, and also in the case of the 6 AXPs and 4 SGRs taken
together (if indeed they can be grouped together as fundamentally the
same kind of objects).

We note in conclusion that the above model would not work as a viable
explanation for AXPs if the underlying neutron stars had
magnetar-like field strengths ($B \sim 10^{14.5}$ G); in this case,
spin-down would be extremely rapid and would produce spin periods of
hundreds of seconds, rather than close to 10 seconds. Thus, it is
unnecessary to invoke neutron stars with non-standard magnetic fields
in the context of this model for AXPs.

\bigskip
\bigskip

We are particularly grateful to Ramesh Narayan for help and advice
during the conception of this paper. We would also like to thank Josh 
Grindlay and Vicki Kaspi for useful discussion.

\begin{figure}[t]
\centerline{\epsfysize=6.0in\epsffile{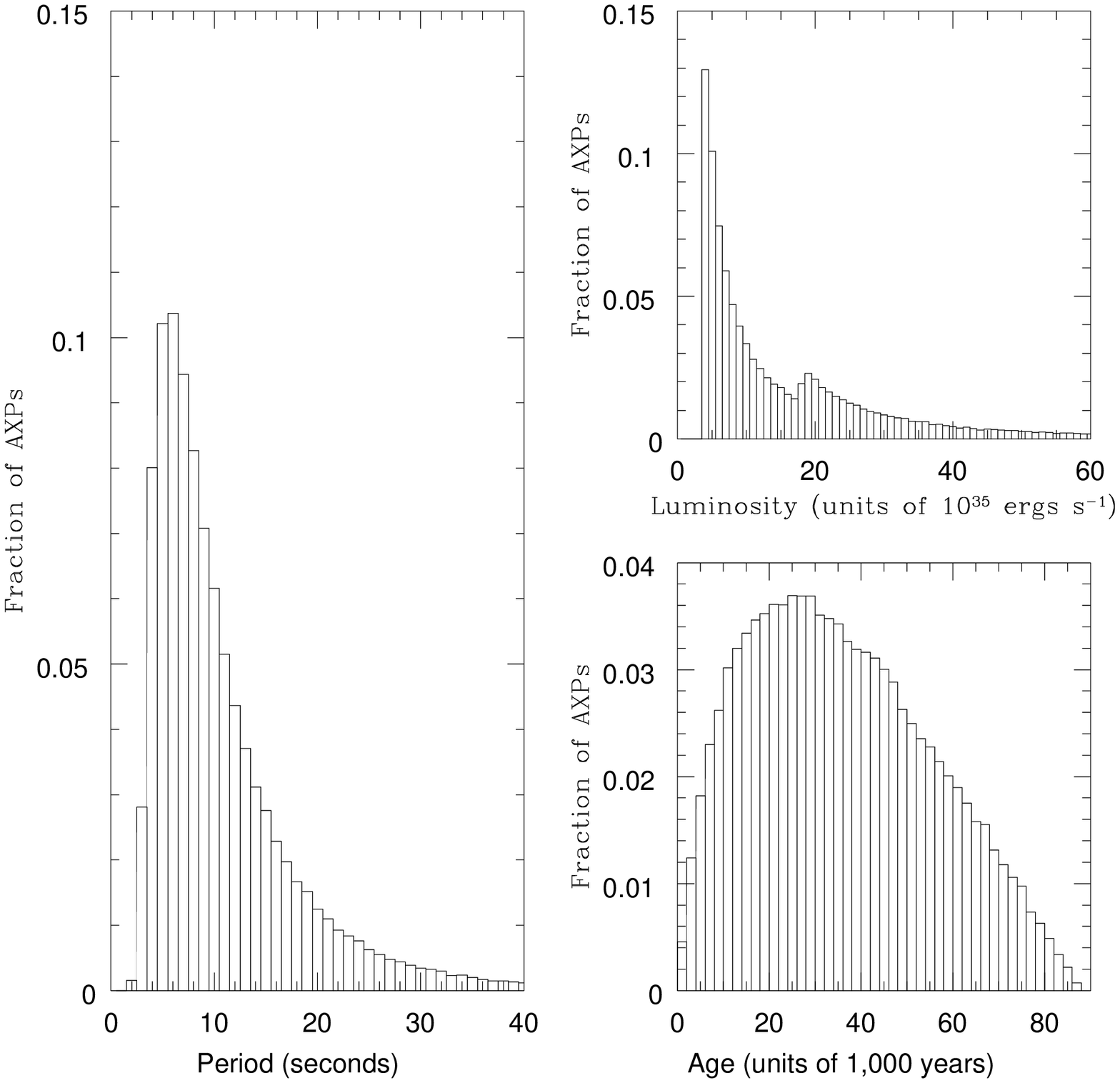}}
\caption{Shown are the distributions of AXP spin period, 
X-ray luminosity and age for a model in which Log$B$ is
distributed normally with a mean of 12.5 and a standard deviation of
0.4, $P_0$ is distributed uniformly between 2 ms and 50 ms, and
Log$(M_d/M_\odot)$ is distributed uniformly between -6 and -1.8. Note
that the luminosity distribution has two peaks; this is due to the
fact that we have chosen a prescription for the X-ray luminosity that is
continuous though not smooth at the time $t=t_{ADAF}$: $L_X \propto 
\dot{M_X}$ for $t<t_{ADAF}$ and $L_X \propto \dot{M_X}^2$ for
$t_{ADAF} \leq t \leq 2 t_{ADAF}$. An AXP is most likely to lie near the
primary peak, which corresponds to low luminosity systems in the 
$t_{ADAF} \leq t \leq 2 t_{ADAF}$ regime, while the region of the
distribution rightwards of the primary peak corresponds to higher
luminosity AXPs in the $t<t_{ADAF}$ regime. Note also that for the
parameters of the neutron star chosen in this paper, the Eddington
luminosity $L_E = 1.8 \times 10^{38}$ ergs $\textrm{s}^{-1}$.}
\label{fig:1}
\end{figure}

\begin{figure}[t]
\centerline{\epsfysize=6.0in\epsffile{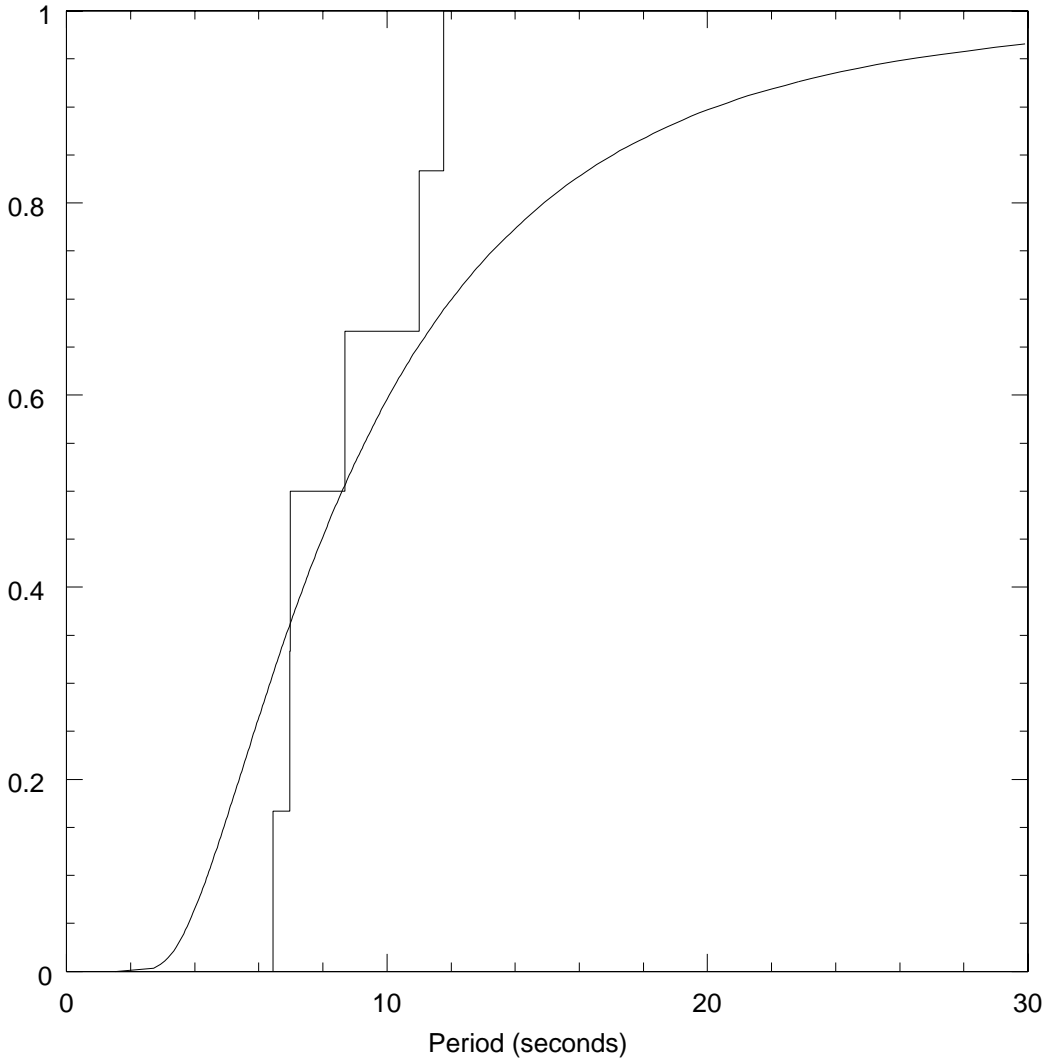}}
\caption{The smooth curve is the cumulative probability distribution 
of the theoretical AXP spin-period distribution for a model 
in which Log$B$ is distributed normally with a mean of 12.5 and a 
standard deviation of 0.4, $P_0$ is distributed uniformly between 2 ms 
and 50 ms, and Log$(M_d/M_\odot)$ is distributed uniformly between -6 
and -1.8. The step-function is the cumulative distribution of the 6
known AXPs. The maximum distance between the two curves is $D=0.31$ 
at a spin period of 11.76 seconds. This value is just significant at a 
Kolmogorov-Smirnov-test level of significance of 0.61.}
\label{fig:2}
\end{figure}

\end{document}